\documentclass[12pt,a4paper]{article}
\usepackage{hyperref}
\usepackage{lscape}
\usepackage{amsmath}
\usepackage{amssymb}
\usepackage{graphicx}
\usepackage{mwe}
\usepackage{tensor}
\usepackage{romannum}
\usepackage{secdot}
\sectiondot{subsection}
\usepackage{subcaption}
\usepackage{hyperref}
\usepackage{enumerate}
\hypersetup{
    colorlinks=true,
    citecolor=blue,
    linkcolor=red,
    filecolor=red,      
    urlcolor=blue,
}
 
\usepackage{doi}
\usepackage{ragged2e}
\begin{document}
	\title{Energy extraction and particle acceleration around a rotating dyonic black hole in $\mathcal{N}=2$, $U(1)^2$ gauged supergravity}

\author{
	{ Anik Rudra$^{a,}$\thanks{rudraanik13@gmail.com}}, 
	{ Hemwati Nandan} $^{b,c,}$\thanks{hnandan@iucaa.ernet.in}, 
	{ Radouane Gannouji$^{d,}$\thanks{radouane.gannouji@pucv.cl}},\\
	{ Soham Chakraborty$^{e,}$\thanks{sohamphy@yahoo.com}}
	{ Arindam Kumar Chatterjee$^{b,}$\thanks{72arindam@gmail.com}}
	\\[0.3cm]
	{\small $^a$ Department of Physics, H. N. B. Garhwal University, S.R.T. Campus, New Tehri - 249199, India.}\\[0cm]
	{\small $^b$ Department of Physics, Gurukul Kangri Vishwavidyalaya, Haridwar - 249407, India.}\\[0cm]
	{\small $^c$ Center for Space Research, North-West University, Mahikeng 2745, South Africa.}\\[0cm]
	{\small $^d$ Instituto de F\'{\i}sica, Pontificia Universidad  Cat\'olica de Valpara\'{\i}so, Casilla - 4950, Valpara\'{\i}so, Chile.}\\[0cm]
	{\small $^e$ Department of Physics, University of York, York YO10 5DD, United Kingdom.}\\[0.2cm]
}
\date{}
\maketitle
\vspace{-1.0cm}
\begin{abstract}
In the present paper, we explore various gravitational aspects such as energy extraction (via the Penrose process and Superradiance), particle collisions around a $\mathcal{N}=2$, $U(1)^2$ dyonic rotating black hole (BH) in the gauged supergravity model. The impact of the rotation parameter ($a$) and the gauge coupling constant ($g$) on the behavior of horizon and ergoregion of the BH is studied. It is of interest to note that, compared with the extremal Kerr BH, the gauge coupling constant, under certain constraints,  can enhance the maximum efficiency of energy extraction by the Penrose process almost double. Under the same constraints, we can extract approximately 60.75\% of the initial mass energy from the BH which is noticeably higher in contrast to the extremal Kerr BH. The limit of energy extraction in terms of the local speeds of the fragments is also examined with the help of the Wald inequality. We identify an upper limit on the gauge coupling constant up to which the phenomenon of Superradiance is likely to occur. Finally, we computed the center-of-mass energy ($E_{CM}$) of two particles with the same rest masses moving in the equatorial plane of the BH. Our study also aims to sensitize $E_{CM}$ to the rotation parameter and the gauge coupling constant for extremal and nonextremal spacetime as well. Especially, for the extremal case, an infinitely large amount of $E_{CM}$ can be achieved closer to the horizon which allows the BH to serve as a more powerful Planck-energy-scale collider as compared to Kerr and any other generalized BHs in the Kerr family explored so far in general relativity. However, $E_{CM}$ for the nonextremal spacetime is shown to be finite and has an upper bound.\\
\vspace{0.1cm}


\end{abstract}

\section{Introduction}

The existence of black holes (BHs) is considered one of the most astonishing consequences of general relativity (GR). One can use BHs to study gravity in strong regime and they can also be used as probes of quantum effects of any fundamental theory, such as supergravity theories. It is thus important to have exact BH solutions derived in the context of supergravity which could be fully analysed by extracting all possible differences with the Kerr black hole (KBH) spacetime in GR. In particular, if a BH is asymptotically AdS and extremal, it plays a crucial role in the AdS/CFT correspondence, principally if the model is embedded in a supergravity model. We prefer $\mathcal{N}=2$ supergravity theory as it has more symmetries than $\mathcal{N}=1$, but at the same time less constrained than higher $\mathcal{N}$ theories (such as the maximal $\mathcal{N}=8$ supergravity). On top of this, phenomenologically, theories with a low number of supersymmetries ( $\mathcal{N}=1, 2$) are closer to the standard model. Chow and  Comp\`{e}re discovered an exact rotating BH solution in 4 dimension with dyonic charges $\mathcal{N}=2$, $U(1)^2$ gauged supergravity in \cite{20} whose geodesic motion has been studied in \cite{21}.

It is worth mentioning that the most powerful conceivable source of energy for galactic nuclei, X-ray binaries, quasars has always been considered of crucial interest in high energy astrophysics. Among several energy extraction mechanisms to explain high energy cosmic events \cite{22},\cite{23}, the Penrose process (particle splitting inside the ergosphere) \cite{24}-\cite{27} and the Blandford–Znajek mechanism (manipulating magnetic field inside the Ergosphere)  \cite{28} are quite famous. The Penrose process has been extensively studied for various spacetimes time and again \cite{29}-\cite{37}. 

Further a similar mechanism known as  superradiance is customarily thought as the {\it wave analogue} of the Penrose process \cite{38},\cite{39}. In this case, a scalar field is boosted while scattering over a BH by extracting energy and angular momentum after reflection off a spinning BH for some range of frequencies \cite{40}. The BH superradiance have been studied from the viewpoint of thermodynamics \cite{41},\cite{42} and BH evaporation as well \cite{43}. The scenario becomes more fascinating if the bosonic waves were repeatedly back onto the BH by a runway process and thus induces a BH bomb \cite{44}-\cite{46}. On the other hand, superradiance could possibly be thought to create exotic particles (axion-like particles) beyond the standard model (SM) in particle physics, leaving an imprint on the dynamics of dark matters. Classically, the event horizon of a black hole is a perfect absorber, so it is no surprise that superradiance can affect directly in black hole geometries. And in case of black holes embedded in a gauged supergravity model due to its rich geometrical structure, this process more effectively extracts energy from the vacuum, even at the classical level in an obvious way.

In view of the above motivation for different energy extraction schemes from a BH, we intend to investigate the dyonic BH emerging in a supergravity model as a particle accelerator. It is now well acquainted by Bañados, Silk, and West (BSW) that the energy of two colliding particles, falling freely from rest outside an extremal Kerr BH, calculated in the center of mass frame, can be arbitrarily high in the limiting case of maximal black hole spin. BSW mechanism has a fertile impact on the viewpoint of ultra high energy collisions as it may generate new interesting physics at Planck-scale (see \cite{47}-\cite{62} for recent developments).

Further, there are several important questions that motivate our analysis: How much insights the solution  \cite{20} can give us in the astrophysical point of view i.e. at what directions can a proliferation of this theoretical model to be tested in the future? What are the functions of gauge coupling with the other regular parameters in generating new physical observations around a spacetime in supergravity? Can we achieve an infinite acceleration for relatively less spinning BH, simply due to the presence of another interesting parameter? What can it teach us about the existence of superradiance for the spacetimes concerned \cite{20} and what are the bounds produced by the parameters of a particular BH spacetime used for investigation? The present paper is aimed to investigate all these question one by one in the background of a $U(1)^2$ dyonic rotating BH spacetime. The paper is organized as follows. We first discuss the horizon structure and ergosphere in detail for the above BH spacetime in Section 2 followed by the geodesic motion in Section 3. Various energy extraction mechanisms are presented in the Sections 4 and 5. The particle acceleration is then discussed in view of Bañados, Silk and West (BSW) mechanism in Section 6. Finally, the results obtained are summarized in Section 7. We have used rescaled units so that the speed of light, the gravitational constant and the BH mass are normalized (8$\pi$$G$ = $c$ = $m$ = 1) throughout the paper except at Section 4 and Section 5 where BH mass is dimensionful.

\section{The structure of spacetime}

The spacetime corresponding to the general solution for $\mathcal{N}=2$, $U(1)^2$ gauged supergravity BH with dyonic charges \cite{20,21}, reads as,

\begin{equation}
ds^2=-\frac{R_g}{B-aA}\left(dt-\frac{A}{\Xi}d\phi\right)^2+\frac{B-aA}{R_g}dr^2+\frac{\Theta_g a^2 sin^2\theta}{B-aA}\left(dt-\frac{B}{a\Xi}d\phi\right)^2+\frac{B-aA}{\Theta_g}d\theta^2\label{12},
\end{equation}

where,\\

$R_g=r^2-2mr+a^2+e^2-N_g^2+g^2 [r^4+(a^2+6N_g^2-2v^2)r^2+3N_g^2(a^2-N_g^2)]$,\\

$\Theta_g=1-a^2g^2cos^2\theta-4a^2N_g cos\theta$, \\

$A=asin^2\theta+4N_g sin^2\frac{\theta}{2}$, \\

$B=r^2+(N_g+a)^2-v^2$, \\

$\Xi=1-4N_gag^2-a^2g^2$. \\

The notations used are as in  \cite{20} \cite{21} and the BH spacetime mentioned above has in general six hairs which are parameterized by mass ($m$), rotation ($a$), electric charges ($e$), magnetic charges ($v$), the gauge coupling constant ($g$) and NUT charge ($N_g$).

\subsection{The Structure of Horizons and Ergosphere}

\begin{itemize}
\item [(i)] The horizon of the spacetime (\ref{12}) satisfies the following equation,

\begin{equation}
r^2-2mr+a^2+e^2-N_g^2+g^2[r^4+(a^2+6N_g^2-2v^2)r^2+3N_g^2(a^2-N_g^2)]=0, \label{eq:horizon}
\end{equation}

As equation (\ref{eq:horizon}) suggests, the BH geometry contains several hairs namely, $m$ (which is sending to unity throughout the section), $a$, $e$, $N_g$ ,$v$ and $g$. In addition to the duality between gravitational mass $m$ and gravitomagnetic mass  $N_g$, there exist another duality between electric charge $e$ and magnetic charge $v$ in this spacetimes. That is why the horizon structure is far different from that is seen for the Kerr BH where the dependence were on the rotation parameter $a$ only. In principle, all the parameters can take all values in $\mathbb{R}$, although not all combinations are physically meaningful. For example if $\Theta_g<0$ the ergoregion goes inside the horizon.  There are instances where singularity can make an appearance outside of the horizon for some combinations $(N_g,v)$. At this stage, the reader should understand that dealing with such a highly nontrivial parameter space is extremely difficult and requires a lot of caution. While many of the parameters in the ungauged theory has been deeply studied and understood in depth, much less is known on the gauged version of it and this is what we are going to focus on.\\
\\Generically, the metric (\ref{12}) admits four horizons which is obvious from the fourth order algebra equation  (\ref{eq:horizon}), including the Cauchy horizon ($r_-$), the event horizon ($r_+$) and the two cosmological horizons since $g$ can be considered as the inverse of AdS scale. Therefore the extremal BH in gauged supergravity can have several types. For first type, the inner horizon and the outer horizon are equal ($r_- = r_+ = r_E$, where $r_E$ is the degenerate horizon). For second type, the outer horizon and one of the cosmological horizons are equal. For third type, the two cosmological horizon  equals and so on. Since only the first of all possible types is of key relevance for our present purpose, we do not pay attention to the rest. From now on we will try to identify only the event horizon and the Cauchy horizon. Due to the extremely intricate nature of the parameter space, we believe that a more exact results of the horizon structure can be found numerically. The key parameters, played much around in our paper i.e. the rotation parameter ($a$) and gauge coupling constant ($g$), are not at all chosen randomly. The values are set from the detailed investigation of how the horizon is affected in the presence of $a$ \& $g$ with the other useful parameters held fixed, furthermore, from the condition of how a black hole may be turned into a nonextremal then extremal and finally becoming a naked singularity solution. The fine-tuning of these parameters remain consistent and as we move deeper into the paper it will become more insightful and intrinsically central to the purpose of our present work. A numerical presentations of the horizon structure (\ref{eq:horizon}) is illustrated in Fig.(\ref{E1}, \ref{E2}) with the extremal value for $r$, $g$ and $a$ taken up to ten decimal places for accuracy. It is observed that if there exist a parameter $\delta^{(g)}$ which measures the difference of radii of the two horizons ( i.e. $\delta^{(g)}=r_+-r_-$) then it decreases with the increase in $g$, as evident from Fig.(\ref{E2})
	 
\begin{figure}[!h]
	\centering
	\begin{subfigure}[b]{0.5\linewidth}
		\includegraphics[width=\linewidth]{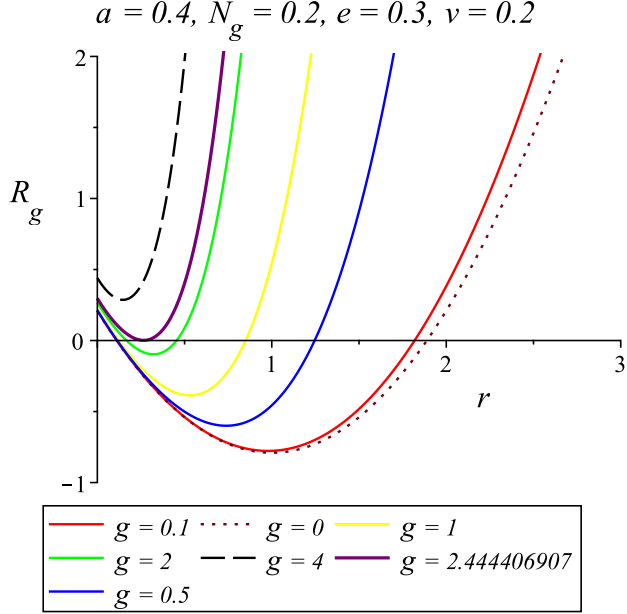}
	\end{subfigure}
	\caption{The structure of the horizon in view of variation of $R_g$ with $r$ for different values of $g$.}
	\label{E1}
\end{figure}

\begin{figure}[!h]
	\centering
	\begin{subfigure}[b]{0.3\linewidth}
		\includegraphics[width=\linewidth]{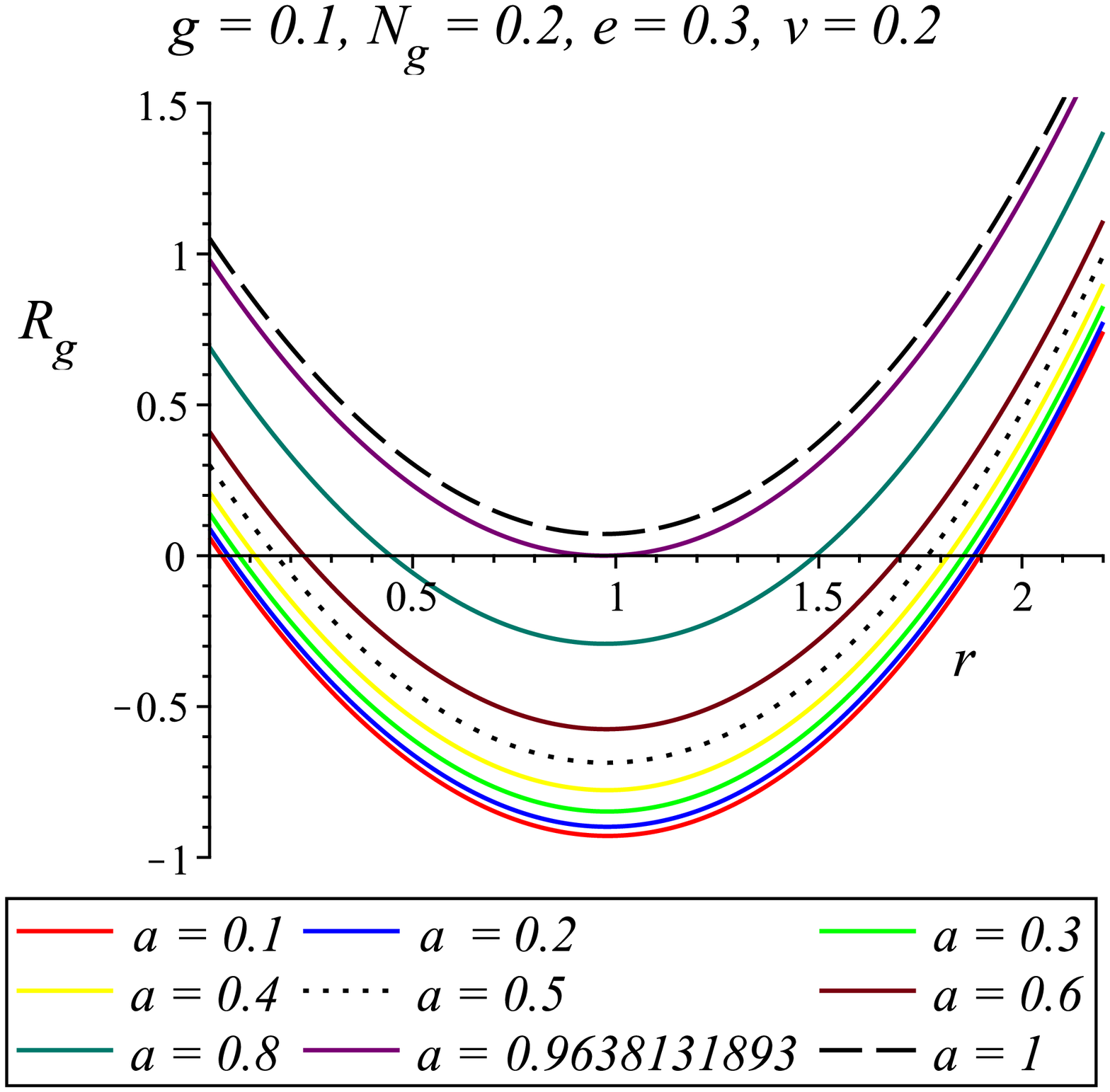}
			\caption {}
	\end{subfigure}
	\begin{subfigure}[b]{0.3\linewidth}
		\includegraphics[width=\linewidth]{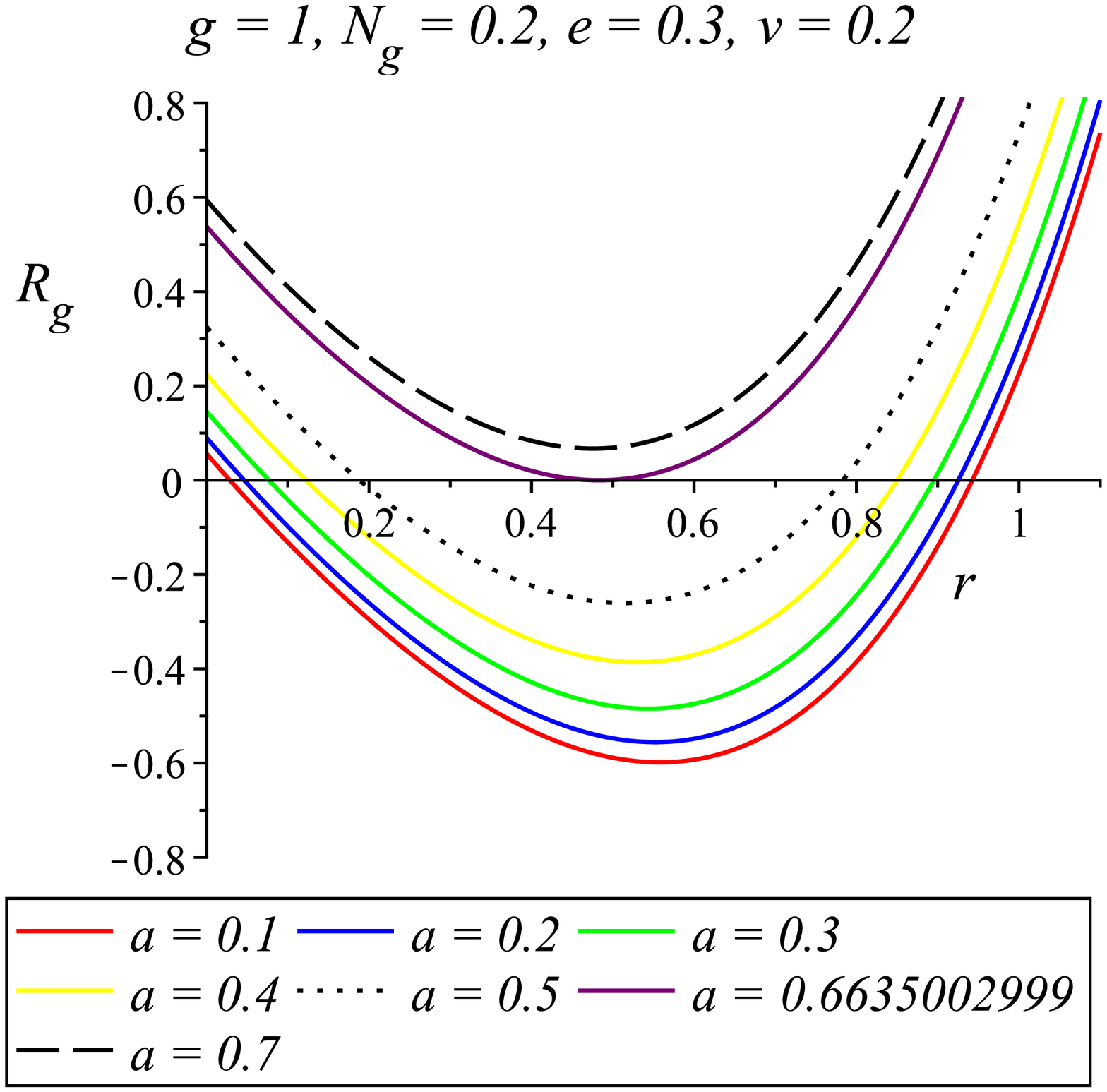}
			\caption {}
	\end{subfigure}
	\caption{The structure of the horizon in view of variation of $R_g$ with $r$ for different values of $a$ at for a fixed value of $g$ with different cases.}
	\label{E2}
\end{figure}

\item [(ii)] One of the important features of the rotating BHs is the existence of the region between the outer horizon and the stationary limit surface (which satisfies $g_{tt}=0$) and by definition, inside this region the asymptotic time translation killing vector becomes spacelike. For the spacetime given by equation (\ref{12}), the static limit surface requires,
\begin{equation}
R_g-\Theta_g a^2 sin^2\theta=0.
\end{equation}
In order to have the ergosphere outside the event horizon, one must impose $\Theta_g>0$, which in turn implies $1-a^2 g^2-4a^2 N_g^2>0$. It is found numerically that the ergosphere depends in a non-trivial way on the parameters of the spacetime. In general, the ergosphere  is found to become smaller when the value of $g$ or $a$ increases as seen from Fig.(\ref{p1}). One can also notice that the structure of the singularity is very complex as shown in \cite{21}. It is not only a ring singularity but for different parameters, this spacetime can have a more complicated three-dimensional structure, which is defined by $r=\sqrt{v^2-(N_g+a\cos\theta)^2}$. The complexity is actually due to the parameters $(N_g,v)$. Even if we have secured the existence of a horizon and an ergosphere exterior to the horizon, the singularity could pop-up outside the horizon. It is therefore always needed to impose an additional condition that the singularity is inside the horizon, which thus requires $r_E>\sqrt{v^2-(N_g+a\cos\theta)^2}$ where $r_E$ is the position of the horizon.
\end{itemize}

\subsection{Comparison to the other kinds of BH in Kerr family}

It should also be emphasized that the BH (\ref{12}) has a number of limiting cases having their own individual implications as discussed extensively in \cite{21}. Moreover, equation (\ref{eq:horizon}) can be re-written as below in the structurally similar form as for Kerr-Newman-AdS BH,
\begin{align}
(1+g^2 r^2)(\alpha+r^2)-2 m r+z=0,
\end{align}
where
\begin{align}
\alpha &= a^2 + 6 N_g^2 - 2 v^2,\\
z &= e^2 - 7 N_g^2 + 2 v^2 + 3 N_g^2 g^2 (a^2 - N_g^2).
\end{align}
These types of quartic equations can always be reduced to a depressed quartic equation which can be solved by Ferrari's method and following the same,  one can define a critical mass,
\begin{align}
m_c=\frac{1}{3\sqrt{6}g}\Bigl(2+2\alpha g^2+\sqrt{x}\Bigr)\Bigl(\sqrt{x}-1-\alpha g^2\Bigr)^{1/2},
\label{eq:critical}
\end{align}
here $ x= 12 g^2 (\alpha+z)+(1+g^2 \alpha)^2$. A study of the positive zeros of the function $R_g$ shows that the line element (\ref{12}) describes a naked singularity for $m<m_c$ and a BH with an outer event horizon and an inner Cauchy horizon for $m>m_c$. Finally, $m = m_c$ represents an extremal BH spacetime. One can notice an important difference with the case of Kerr-Newman-AdS BH for which $\alpha=a^2\geq 0$ and $z=e^2\geq 0$. In Kerr-Newman-AdS BH case, $m_c>0$ and therefore we have a minimum mass of the BH, while the situation for the metric (\ref{12}) is different because of undefined signature of $A$ and $z$.  One can notice that the condition (\ref{eq:critical}) reduces in the ungauged case ($g=0$) to $m_c=\sqrt{a^2+e^2-N_g^2}$ which is similar to the Kerr-Newman-Taub-NUT BH condition.

\section{The Geodesic Motion}

Let us consider a motion for a particle with mass $m_0$ falling in the background of a $\mathcal{N}=2$ rotating dyonic gauged supergravity BH. Now it is straightforward to see that for this BH spacetimes, the Hamilton- Jacobi equation for the equatorial plane

\begin{equation}
\frac{1}{2} g^{\mu\nu} \frac{\partial S}{\partial x^\mu} \frac{\partial S}{\partial x^\nu}=-\frac{\partial S}{\partial \lambda}
\label{eq:HJ}
\end{equation}
is separable and yields for each coordinate a corresponding first order differential equation

\begin{equation}
\frac{{\rm d}t}{{\rm d}\tau}=
\frac{B ( E B-a L \Xi) }{R_g}
+\frac{A (L \Xi - E A)}{\Theta_g\sin^2\theta}
\label{eq:t_eqn}
\end{equation}

\begin{equation}
\frac{{\rm d}\phi}{{\rm d}\tau}=
\frac{a\Xi ( E B-a L \Xi) }{R_g}
+\frac{\Xi (L \Xi - E A)}{\Theta_g\sin^2\theta}
\label{eq:phi_eqn}
\end{equation}

Here we used  the normalization relation $g_{\mu \nu}u^\mu u^\nu=(-1,0)m_0^2$  and defined constants $m_0$, $E$ and $L$ which corresponds to rest mass, conserved and axial part of the angular momentum of the particle respectively. These constants are related via $m_0^2=-p_\mu p^\mu$, $E=-p_t$ and $L=p_\phi$ (with $c=1$ and $G=1$)

\begin{figure}
\begin{centering}
\includegraphics[scale=.7]{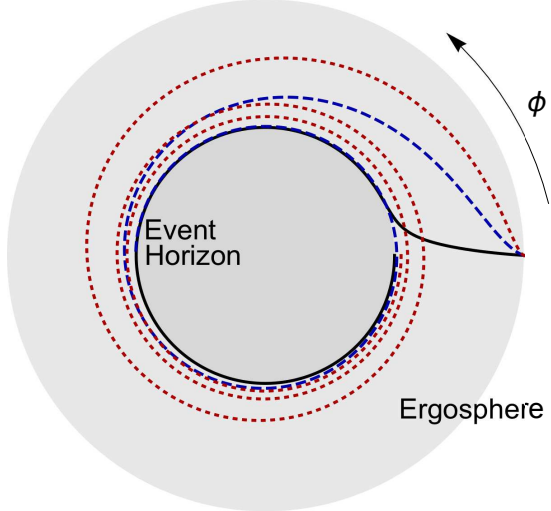}
\par\end{centering}
\caption{Various orbits from the ergosphere to the event horizon for differnt values of $g$ ($g=0.5$ in black, $g=1$ in dashed blue and $g=2$ in red dotted), while other parameters are fixed to 1 except $e=0.1$. For the different values of $g$, the ergosphere and the event horizon have been rescaled to coincide.}  
\label{p1}
\end{figure}

Further the equation involving the conjugate momenta $p_r$ can be written in the following form
\begin{equation}
(B-aA)\dot{r}=\pm \sqrt{\mathcal{R}},
\end{equation}
where,

\begin{equation}
\mathcal{R} \equiv \mathcal{P}^2-R_g(K+m_0^2B),
\end{equation}
and

\begin{equation}
K=-am_0^2(a+2N_g)+(E(a+2N_g)-L \Xi)^2
\end{equation}
and

\begin{equation}
\mathcal{P}^2 \equiv (EB-\Xi a L)^2.
\end{equation}
With the so-called Mino time as $d\lambda=(B-aA)d\tau$, we have,
\begin{align}
\frac{{\rm d}r}{{\rm d}\tau}=\pm \sqrt{\mathcal{R}}
\label{eq:rdot}
\end{align}
from which one can obtain ${\rm d}r/{\rm d}\phi$ and integrate it. It is instructive to pay attention in the Fig.(\ref{p1}) that orbits are falling faster into the BH when $g$ is smaller. Which  is consistent because $g$ plays the same role that the inverse of the AdS scale which produces a repulsive effect.\\

Clearly, $(B-aA)$ serves here as a prefactor in $r$, $t$ and $\phi$ motion of this BH spacetime. It exactly coincides with the prefactor $\Sigma=r^2$ on the equatorial plane for the Kerr BH in GR when $N_g=v=e=g=0$ \cite{25}.\\

\section{Energy Extraction: Penrose Process}

 The existence of an ergosphere was first pointed out by Penrose in 1969 \cite{24} which provides a way to extract energy from a rotating BH by sending a test particle to the ergosphere where it decays into two identical particles at a turning point, $\dot r=0$, in its geodesic trajectory: one with positive energy which escapes the BH and the other with negative energy absorbed by the BH.
   It is thus important to find the limits on the energy which a particle at a particular location can have. Here, we will consider the possibilities of energy extraction from a BH being used by us and immediately  afterward, we will discuss the original Penrose process along with the study of the bounds from Wald inequality.\\

From eq.(\ref{eq:rdot}), with the condition $\dot r=0$, we have $\mathcal{R}=0$ which leads to, 
\begin{equation}
E=\frac{aB-(a+2N_g)R_g\pm \sqrt{R_g(r^2+N_g^2-v^2)}\sqrt{1+m_0^2 \Delta}}{\Delta},
\end{equation}  
where
\begin{align}
\Delta=\frac{B^2-R_g(a+2N_g)^2}{L\Xi},
\end{align}
or alternatively,
\begin{align}
L=E\frac{aB-(a+2N_g)R_g\pm(r^2+N_g^2-v^2)\sqrt{R_g}\Bigl(1+\frac{m_0^2}{E^2}\frac{a^2-R_g^2}{r^2+N_g^2-v^2}\Bigr)^{1/2}}{\Xi (a^2-R_g)}.\label{el}
\end{align}
The '$\mp$' signs in eq. (\ref{el}) corresponds to co-rotating and counter rotating orbits respectively. In order to have positive energy in the Kerr limit, we must retain only the positive sign. We can also see easily that a necessary condition for negative energy is $L<0$ which means only counter rotating particles can possesses negative energy.

In energy extraction process, the incident massive particle with $(E^{(0)},L^{(0)})$ breaks up into two massless particles with energy and angular momentum: $(E^{(1)},L^{(1)})$ for a particle falling into the BH and $(E^{(2)},L^{(2)})$ for the particle leaving the ergosphere. 
We consider total energy and angular momentum as conserved at the point of break which then reads, 
\begin{align}
E^{(0)}&=E^{(1)}+E^{(2)},\\
L^{(0)}&=L^{(1)}+L^{(2)}.
\end{align}

The efficiency of energy extraction takes the following form,
\begin{equation}
\eta=-\frac{E^{(1)}}{E^{(0)}}=\frac{1}{2} \Bigl(\sqrt{1+\frac{a^2-R_g^2}{(E^{(0)})^2(r^2+N_g^2-v^2)}}-1\Bigr).
\end{equation}
Further when the incident particle splits at the horizon $r_E$, one can obtain the maximum efficiency as below,
\begin{equation}
\eta_{max}=\frac{1}{2} \left[\sqrt{1+\frac{a^2}{(E^{(0)})^2(r_E^2+N_g^2-v^2)}}-1 \right].
\label{eq:eta}
\end{equation}
It is interesting to note that this result do not depend explicitly on the parameters $(g,e,m)$. Of course these parameters appear in the equation for the horizon, $r_E$. One can easily recover the standard results for KBH case with $N_g=v=e=g=0$ for which $\eta_{\text{max}}=0.207$ \cite{25}.
\begin{figure}
\begin{centering}
\includegraphics[scale=.7]{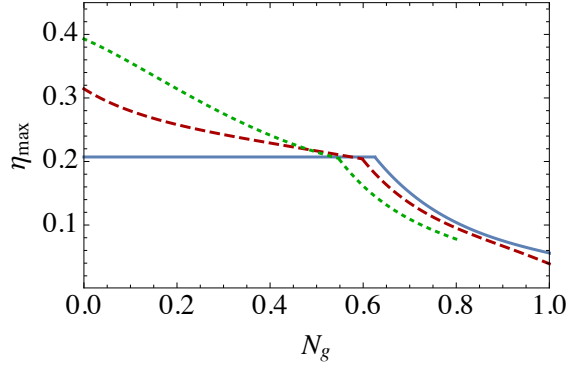}
\par\end{centering}
\caption{Efficiency of the Penrose process as a function of  $N_g$ for different values of g (i.e. $g=0$ in blue, $g=0.5$ in dashed red and $g=1$ in dotted green). Here $v=0$ and the mass is fixed to be the critical mass i.e. $m=m_c$ and  $E^{(0)}=1$.
}  
\label{p2}
\end{figure}

We see in Fig. (\ref{p2}) that efficiency is similar to  KBH for $g=0$ and for $N_g<0.65$ and smaller for $N_g>0.65$. One can therefore conclude that Kerr-Newman-NUT spacetime is less efficient than its Kerr counterpart as already discussed in \cite{60}. One can also notice that for larger $g$, it has better efficiency.

In table (\ref{table2}) we have looked to efficiency as a function of $a$, in the very particular case where $N_g=v$ for an extremal BH. In this case, the formula (\ref{eq:eta}) reduces to the Kerr formula. But because the horizon can have different values compared to Kerr spacetime, the efficiency will be different.

\begin{table}[!h]
	\centering
	\begin{tabular}{|c|c|c|c|c|}
		\hline
		{\bf $g$} & {\bf $a$} & {\bf $r_E$} & {\bf $\eta$} \\
		\hline
		0.1 & 0.9638131893 & 0.9711 & 0.2044 \\
		\hline
		0.5 & 0.8235924578 & 0.6096 & 0.2791\\
		\hline
		1 & 0.6635002999 & 0.4835 & 0.3489 \\
		\hline
		2 & 0.4600967101 & 0.3080 & 0.3988 \\
		\hline
	\end{tabular} 
		\caption{The  efficiency of the energy extraction via Penrose process in the nonextremal BH case for different values of $g$ and $a$ with $e=0.3$ and $N_g=v=0.2$.}
		\label{table2}
\end{table}

One may note that the Penrose process can be enhanced as the parameter $g$ increases and decreases with $N_g$.

However, the question on how much energy could maximally be extracted not from a particle orbiting towards the rotating BH but from the BH itself is answered in the next subsection.

\subsection{Extraction of the initial mass energy}

A series of Penrose mechanism could extract all the angular momentum of a Kerr BH, leaving an ordinary  Schwarzschild BH. No further energy can then be extracted from the resulting  black hole (except by quantum Hawking processes). The mass of the BH is an observer dependent quantity.  As the horizon gets closer, the BH mass increases. The  mass of the BH of its final fate is termed as 'irreducible mass' ($m_{irr}$) which can't be decreased by any classical process. According to the new horizon mass theorem \cite{64}, the horizon mass is always twice of the irreducible mass observed at infinity. When the BH solution is extremal, the irreducible mass can be related to the horizon mass ($m$) in the following way,
\begin{equation}
m_{irr}=\sqrt{\frac{1}{2}mr_+}.
\end{equation}
The maximum amount of energy that can therefore be extracted from an extremal $U(1)^2$ supergravity BH is given by,
\begin{equation}
m-m_{irr}=m-\sqrt{\frac{mr_+}{2}}.
\end{equation}
For extreme strongly coupled but less spinning supergravity BH, under the constraint $N_g=v$, it is possible to extract approximately 60.7 percent of the total energy while it is only 29.3 percent for the extreme KBH as mentioned in table (\ref{table3}). Although the amount of energy extraction increases with the increase in $g$, but $g$  has an upper limit (as indicated earlier).
\begin{table}[!h]
	\centering
	\begin{tabular}{|c|c|c|c|}
		\hline
		{\bf $g$} & {\bf $r_E$} & {\bf $m-m_{irr}$}\\
		\hline
		0.1 & 0.9711 & 0.3031 \\
		\hline
		0.5 & 0.6906 & 0.4123 \\
		\hline
		1 & 0.4835 & 0.5083 \\
		\hline
		2 & 0.3080 & 0.6075 \\
		\hline
	\end{tabular} 
		\caption{The net extracted energy $m-m_{irr}$ from the extremal BH for different values of $g$, where $m_{irr}$ is the irreducible mass of the BH.}
		\label{table3}
\end{table}

\subsection{The Wald Inequility}

The Wald inequility \cite{65} further establishes lower bounds on the local speeds of the fragments in order  the Penrose process to take place. It explains the origin and limitations of the Penrose process depending on the geometry of the spacetime as well as the velocity components of the fragments. Here the detailed derivation of Wald inequility is not presented as it is almost a parallel treatment to that of the KBH case developed in \cite{25} and we to follow the Ref. \cite{25}. Let us imagine, a particle with four-velocity $U^\mu$ and conserved energy $\tilde{E}$ that splits up and emits a fragment with energy $\tilde{E}'$ and four-velocity $u^\mu$. Now the Wald inequility imposing the limits on $\tilde{E}'$, gives three velocities of the fragment $\vec{v}$ as measured in the rest frame of the incident body becomes,

\begin{equation}
\gamma \tilde{E}-\gamma v (\tilde{E}^2+g_{tt})^{\frac{1}{2}} \le \tilde{E}' \le \gamma \tilde{E}+\gamma v(\tilde{E}^2+g_{tt})^{\frac{1}{2}},
\end{equation}

where $\gamma$ is the Lorentz factor, i.e. $\gamma=\frac{1}{\sqrt{1-v^2}}$. For $\tilde{E}'$ to be negative, we must have for our spacetime at $\vartheta=\frac{\pi}{2}$ and on the horizon,

\begin{equation}
|v|>\frac{1}{\sqrt{1+\tilde{E}\frac{a^2}{r_+^2+(N_g^2+a)^2-v^2}}}.
\end{equation}

Before any extraction of energy, the fragments must possess relativistic energies as evident from Table (\ref{table4}). We can consider the extreme cases as well. For the extreme Kerr spacetime $|v|>0.707$, as derived in \cite{25}, considering the customary value $\tilde{E}=1, 2$, it is observed that $|v|$ decreases as the gauge-coupling constant $g$ becomes stronger. 

\begin{table}[!h]
	\centering
	\begin{tabular}{|c|c|c|c|}
		\hline
		{\bf $g$} & {\bf $a_E$} & {\bf $r_E$} & {\bf $|v|$}\\
		\hline
		0.1 & 0.9638131893 & 0.9711 & 0.8417\\
		\hline
		0.5 & 0.8235924578 & 0.6096 & 0.8284\\
		\hline
		1.0 & 0.6635002999 & 0.4835 & 0.8251\\
		\hline
	\end{tabular} 
		\caption{Lower bounds on the local speed ($|v|$) of fragments for different values of $g$ and $a$ in the extremal BH.}
		\label{table4}
\end{table}

\section{Superradiance}

The Penrose mechanism adheres only about a possibility that allows for a particle to come out of a rotating BH with more energy than its 'parent particle'. Practically, Penrose processes are not likely to be important in astrophysics because the required conditions can not easily realized. A more general situation could occur where one has to put some medium or some matter field in some background spacetime that provides the arena for superradiance because superradiance requires dissipation. It is thus important to remember that superradiance may occur in vacuum provided the given spacetime is curved. As compared to Penrose process, superradiance is analogous effect for the waves. A part of the wave is absorbed while it reaches to BH 
and a part of the wave is reflected. In some cases the absorbed wave carries negative energy while the reflected wave is amplified. We explore here the superradiant scattering of radiation by an $U(1)^2$ dyonic rotating spacetime. Let us start with current continuity equation,

\begin{equation}
\Box \Phi= \frac{1}{\sqrt{-g}} \partial_\mu ( \sqrt{-g}g^{\mu\nu} \partial_\nu \Phi),
\end{equation} 

which is associated with the energy flux vector field $\Phi$ (i.e. a test field). Considering a simple wave mode of frequency $\omega$,

\begin{equation}
\Phi=e^{-i\omega t}e^{im\phi} \vartheta(\theta)R(r).
\end{equation} 

 We closely follow the derivations for superradiance as presented in \cite{60} and the expression obtained for the energy flux lost per unit time (power) is given below,

\begin{equation}
dP=\omega(\omega-m \Omega_H) \left(\frac{B}{B-aA}\right)_{r_E} \iint (B-aA)_{r_E} \Theta(\theta)^2 sin^2 \theta d\theta d\phi,
\end{equation}

\begin{equation}
P \sim \omega(\omega-m\Omega_H)[r_E^2+(N_g+a)^2-v^2]=constant. \label{13}
\end{equation}

where $\Omega_H=\frac{a\Xi}{B}$ is the angular velocity of the outer horizon.\\

If $\omega>m \Omega_H$ $P$ is positive, then the superradiance is not possible. However, on the other hand, the superradiance occurs if $\omega$ lies in the range $0<\omega<m\Omega_H$. Within this inequality range, it is evident from equation (\ref{13}) that a wave mode is amplified indeed by the BH. The angular momentum quantum number ($m$) must be non-zero as it has to take away angular momentum from the BH. The value of $\Omega_H$ is important and for extremal $U(1)^2$ supergravity spacetime with constraint $N_g=v$ the table (\ref{table5}) clearly indicates that $\Omega_H$ essentially changes sign as $g$ becomes larger. So in order to have superradiance $g$ must not exceed 1 in the table (\ref{table5}) below, that makes the frequency range smaller. The prefactor $[r_E^2(N_g+a)^2-v^2]$ increases and decreases with respect to $N_g$ and $v$ respectively. The prefactor is related in modifying the magnitude of the amplification. 

\begin{table}[!h]
	\centering
	\begin{tabular}{|c|c|c|c|c|c|}
		\hline
		{\bf $g$} & {\bf $\Xi$} & {\bf $r_E$} & {\bf $a$} & {\bf $\Omega_H=\frac{a\Xi}{B}$} \\
		\hline
		0.1 & 0.9136 & 0.9711 & 0.9638131893 & 0.3900 \\
		\hline
		0.5 & 0.6657 & 0.6906 & 0.8235924578 & 0.3692 \\
		\hline
		1 & 0.0289 & 0.4835 & 0.6635002999 & 0.0204 \\
		\hline
		2 & -1.31 & 0.3080 & 0.4600967101 & -1.2285 \\
		\hline
	\end{tabular} 
		\caption{The angular velocity of the extremal BH at the horizon for different values of $g$ and $a$.}
		\label{table5}
\end{table} 

\section{CM Energy and Particle Collision}

In this section, the  CM energy of the two particles colliding in the equatorial plane of $U(1)^2$ dyonic rotating BH is investigated. We further assume that the particles with same rest mass $m_0$ but having different energies coming from the infinity has $\frac{E_1}{m_0}=\frac{E_2}{m_0}=1$ where they were initially at rest. Finally, the particles are approaching towards the event horizon of the BH (mentioned above) with different angular momenta $L_1$ and $L_2$. The general form of CM energy of two colliding particles $ i (i=1,2) $ \cite{47} is given by,

\begin{equation}
\frac{E_{CM}^2}{2m_0^2}=1-g_{\mu\nu} u_{(1)}^a u_{(2)}^b
\end{equation}
where, $u_{(1)}^a$ and $u_{(2)}^b$ are the four velocities of two particles respectively. For the spacetime considered here the above formula reads as,
\begin{multline}
\frac{E_{CM}^2}{2m_0^2}=\frac{1}{R_g(B-aA)}[R_g (B-aA)+(B^2-R_g A^2)-(R_g-a^2)\Xi^2 L_1 L_2 \\ -(aB-R_gA)\Xi(L_1+L_2)-
\sqrt{(B-\Xi a L_1)^2-R_g(A-\Xi L_1)^2-R_g B} \\ \sqrt{(B-\Xi a L_2)^2-R_g(A-\Xi L_2)^2-R_g B}], \label{11}
\end{multline}
which is similar to the KBH in GR \cite{47} for $g=0$, $N_g=0$, $e=0$ and $v=0$.\\

Since $E_{CM}$ is an invariant scalar, it serves as an observable, therefore independent of co-ordinate choices. This ensures the validity of the formula given in equation (\ref{11}) in Special Relativity as well as GR with $m_0$ = 1, $m$ = 1.

\subsection{Near-Horizon Collision in Extremal Spacetime}

\begin{itemize}
	\item It is worthwhile to study the properties of the collisional energy (\ref{11}) as the radius $r$  approches to the horizon $r_E$ of an extremal supergravity black hole. However, a particle with very large angular momentum can't reach the horizon if it falls freely from rest at infinity. So there must be a range for the angular momentum to ensure that only the particle with critical angular momentum ($L=L_c$) can reach the horizon of the BH. And for this we  are going to use the effective potential method. The effective potential for a timelike particle moving along the geodesic in the equitorial plane is
		\begin{equation}
		V_{eff}=-\frac{1}{2}(\dot{r})^2
		\end{equation}
		 which approaches $0$ at infinity. So a condition for the particle falling freely from rest at infinity to reach the horizon is $V_{eff}\le 0$ for any positive $r$. The full expression of effective potential of the test particle can be obtained by just looking at the equation (\ref{eq:rdot}). The physical particles must satisfy the forward-in-time condition which gurantees $\frac{dt}{d\tau} \ge 0$. At $r \to r_E$, this condition leads to the following relation
	
	\begin{equation}
	E-\Omega_H L \ge 0 \label{3},
	\end{equation}
	where, $\Omega_H=\frac{a\Xi}{B}$ is the angular velocity of the extremal BH at the horizon. The critical angular momentum is given by $L_c=\frac{E}{\Omega_H}$. 
	The numerical data for the critical values of angular momenta are listed in table (\ref{table6}). 
	
	\begin{table}[!h]
		\centering
		\begin{tabular}{|c|c|c|c|c|}
			\hline
			{\bf $g$} & {\bf $r_E$} & {\bf $a$} & {\bf $L_c=\frac{B (r_E)}{a\Xi}$} \\
			\hline
			0.1 & 0.9711 & 0.9638131893 & 2.5640\\
			\hline
			1 & 0.4835 & 0.6635002999 & 48.990\\
			\hline
			2.444406907 & 0.2660 & 0.4 & -0.5252\\
			\hline
		\end{tabular} 
		\caption{Numerical evaluation of critical angular momentum ($L_c$) for an extremal BH for different values of $g$ and $a$.}
		\label{table6}
	\end{table} 
Now it is easy to verify that if one of the particle's angular momentum is in a proper range, the effective potential is negative, otherwise it is positive near the horizon and the particle can not reach the horizon as is clearly indicated from Fig.(\ref{L}).

\begin{figure}
	\begin{centering}
		\includegraphics[scale=.4]{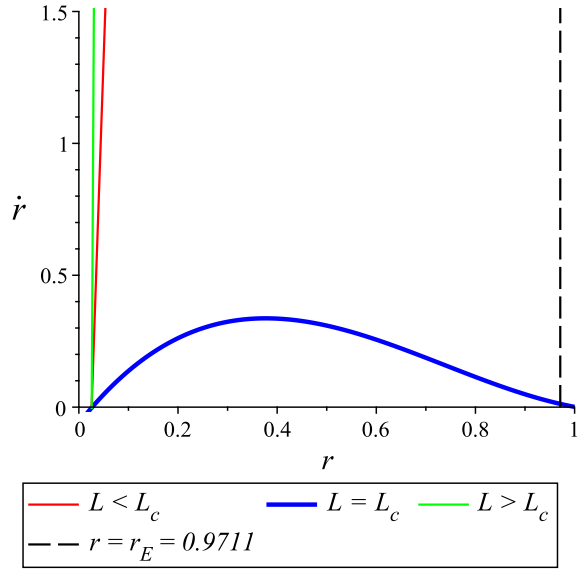}
		\par\end{centering}
	\caption{The variation of $\dot{r}$ with $r$ for different values of angular momentum $L$ has been shown for the extremal BH with $g = 0.1$. The bold blue line corresponds to the value $L_c = 2.5$ allowing the particle to reach the horizon.} 
	\label{L}
\end{figure}

 \item We now analyze the CM energy formulated in equation (\ref{11}) of two colliding particles as $r \to r_E$ for the extremal gauged supergravity BH. The limit $r \to r_E$ makes equation (\ref{11}) in indeterminate form. By applying L'H$\hat{\text{o}}$pital's rule, one can easily calculate the limiting value of $E_{CM}$ (as $r \to r_E$) with critical angular momentum. Naively, from Fig. (\ref{FigCM1}, \ref{FigCM2}, \ref{FigCM3}) $E_{CM}$ diverges at the horizon. Compared with the result for the Kerr BH \cite{47}, here the spin of the BH may deviate from its maximum value but we can nonetheless obtain an arbitrary high center of mass energy due to the commanding presence of gauge coupling  constant. In particular, the interplay between the rotation parameter and the gauge coupling constant is very interesting at this point because the result of achieving infinitely high $E_{CM}$ does not change in the case of BH with less spinning, strongly coupled or vice versa.\\
	

The unbounded nature of $E_{CM}$ reveals the high energetic particle collisions near the event horizon of the extremal BH. This in turn, establishes the fact that extremal $U(1)^2$ dyonic rotating BH  can be a particle accelerator to generate physics at high energy scale.

\begin{figure}[!h]
	\centering
	\begin{subfigure}[b]{0.375\linewidth}
		\includegraphics[width=\linewidth]{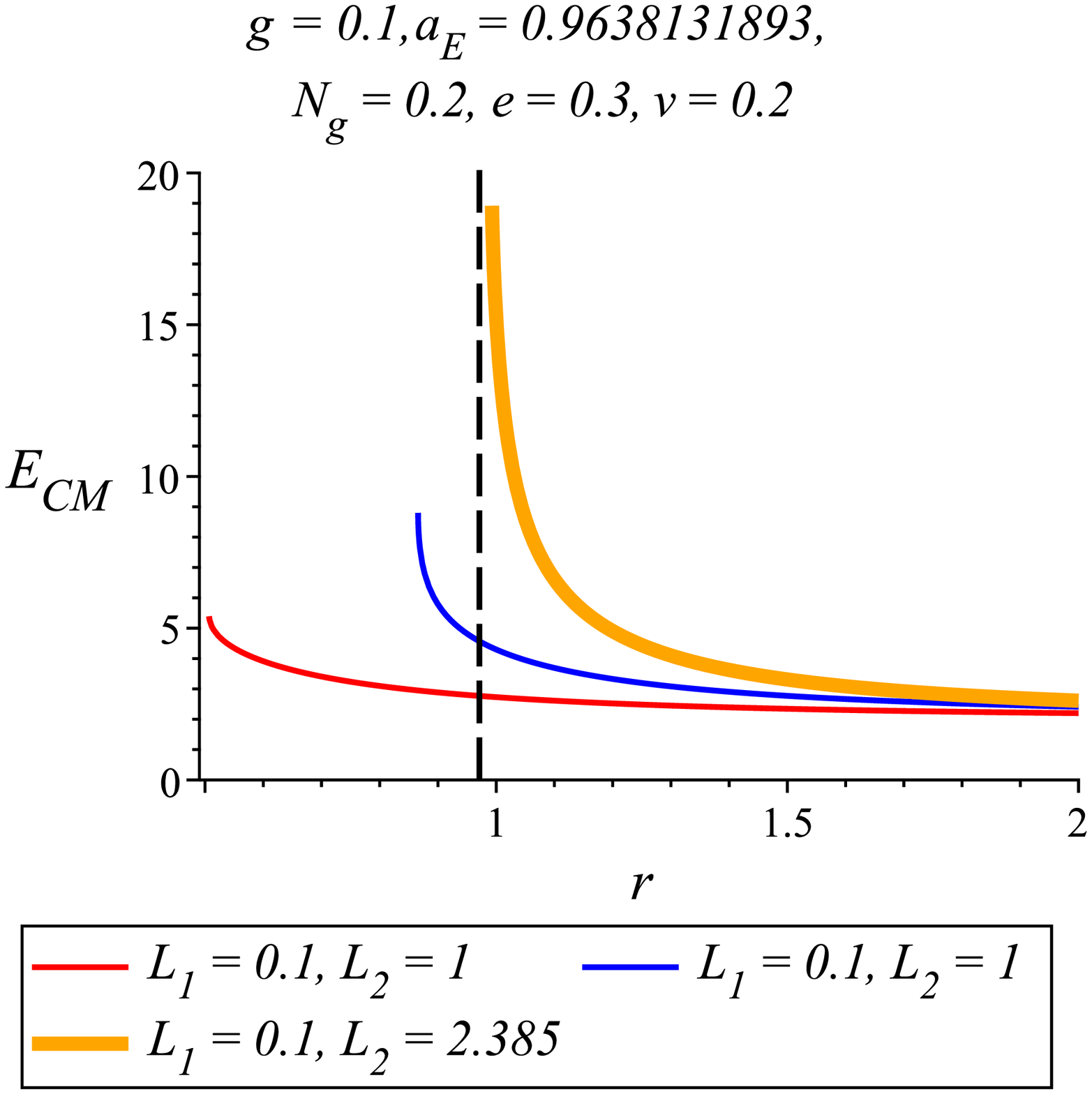}
		\caption{(i)}
	\end{subfigure}
	\begin{subfigure}[b]{0.375\linewidth}
		\includegraphics[width=\linewidth]{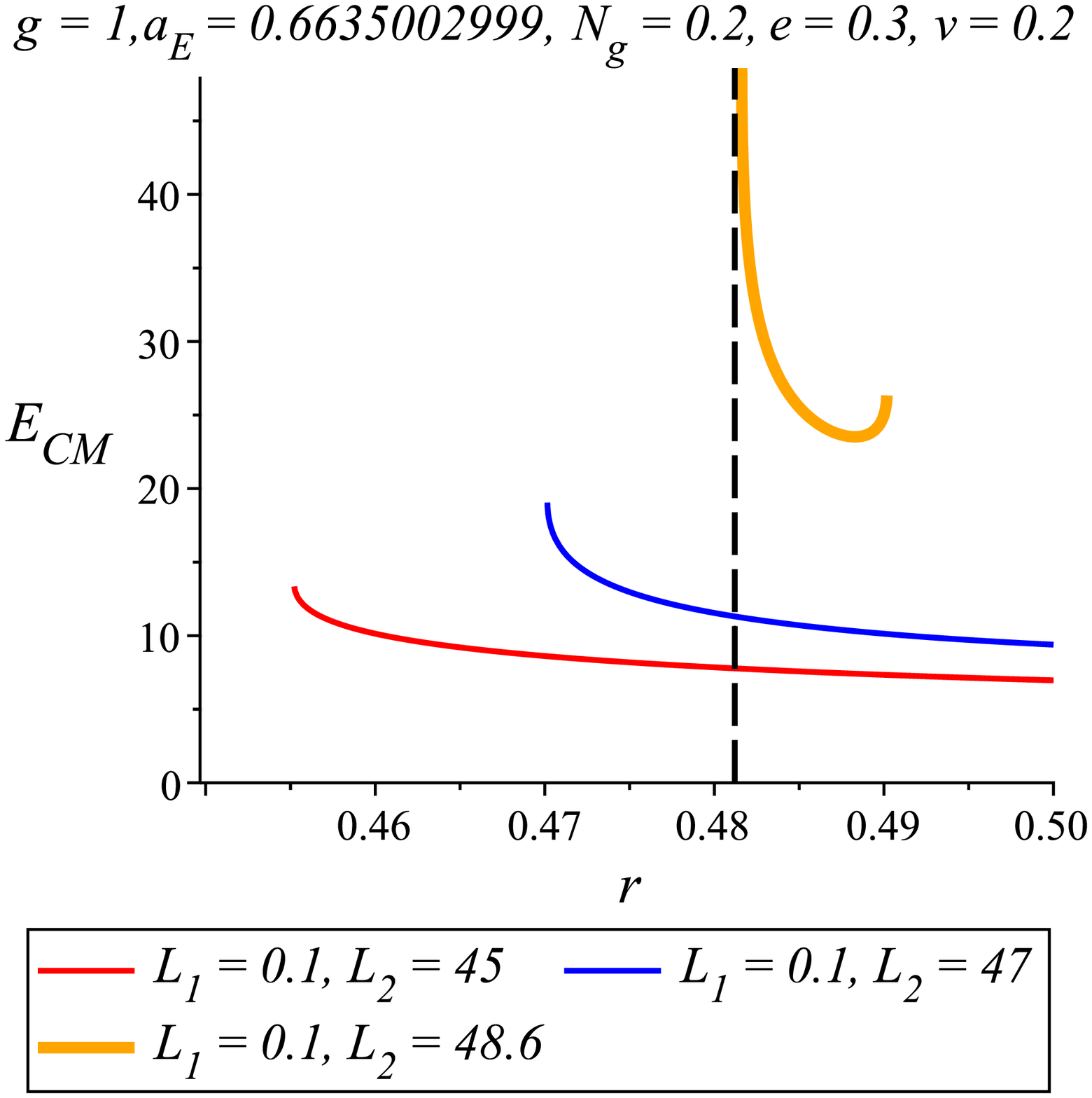}
		\caption{(ii)}
	\end{subfigure}
	\caption{For an extremal BH case the variation of $E_{CM}$ with $r$ for three combinations of $L_1=L_c$ and $L_2$,  where the vertical black dashed line corresponds to the degenerate horizon.}
	\label{FigCM1}
\end{figure}

\begin{figure}[!h]
	\centering
	\begin{subfigure}[b]{0.375\linewidth}
		\includegraphics[width=\linewidth]{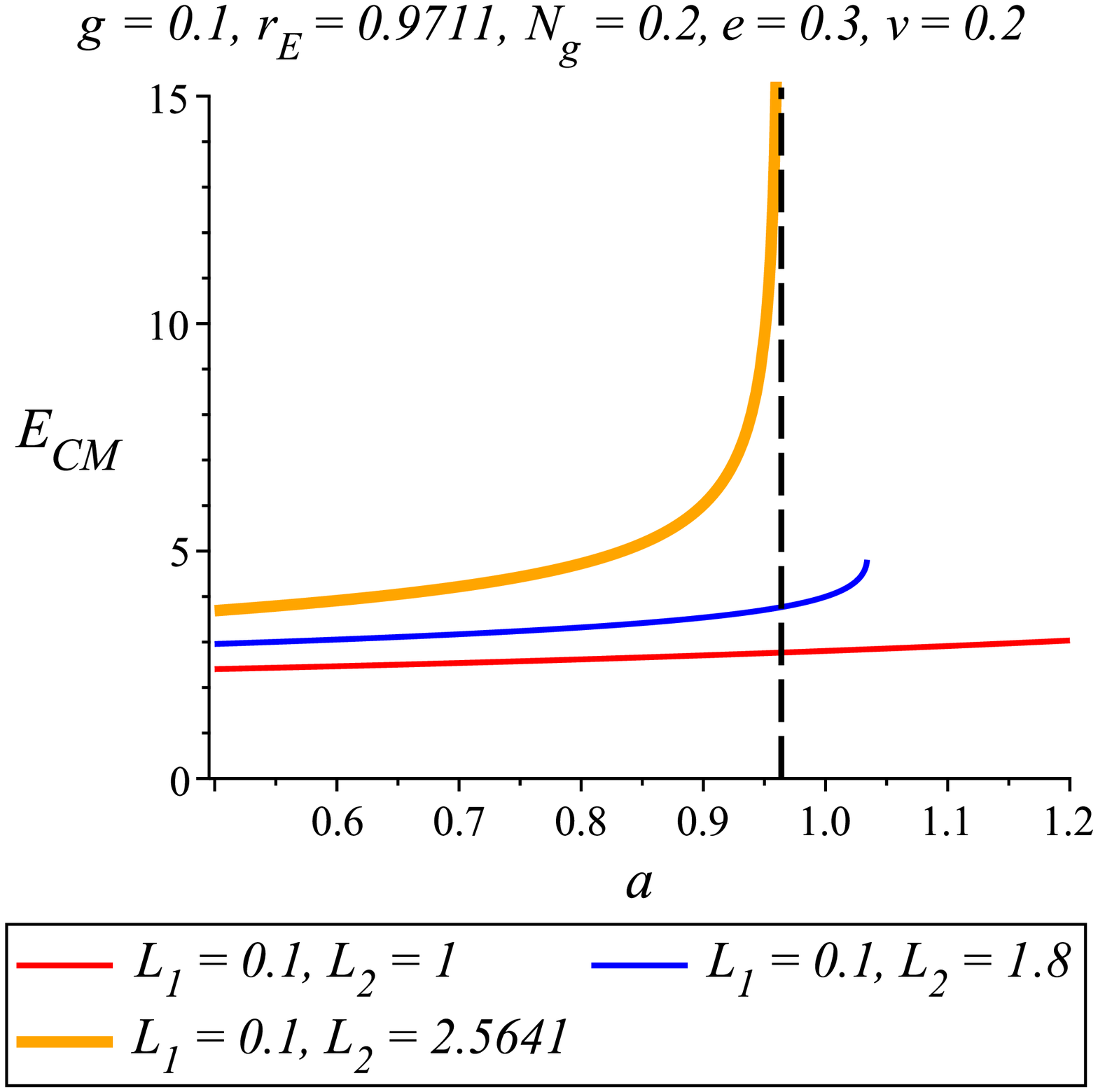}
		\caption{(i)}
	\end{subfigure}
	\begin{subfigure}[b]{0.375\linewidth}
		\includegraphics[width=\linewidth]{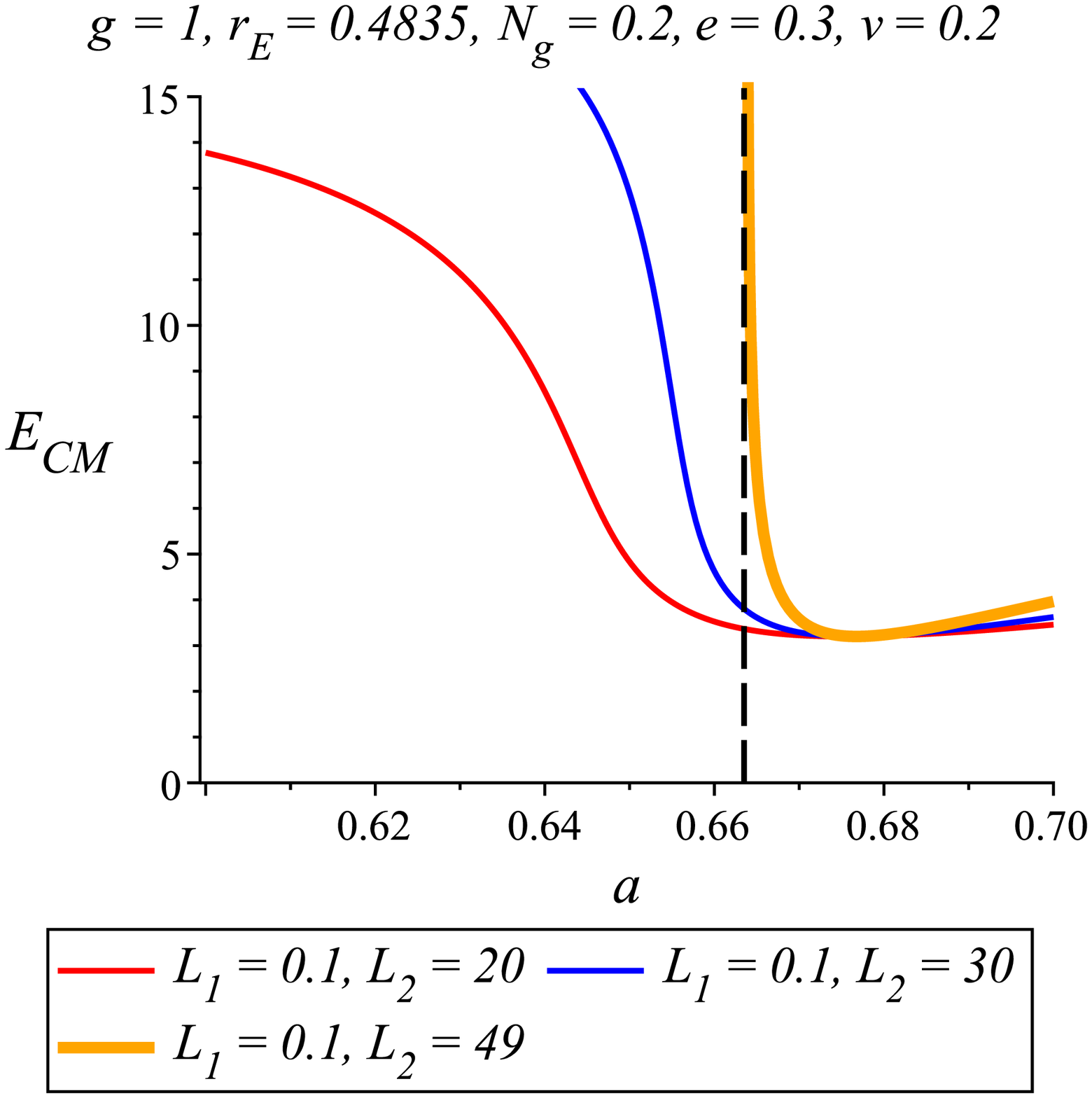}
		\caption{(ii)}
	\end{subfigure}
	\caption{For an extremal BH case the variation of $E_{CM}$ with $a$ for three combinations of $L_1=L_c$ and $L_2$, where the vertical black dashed line corresponds to the degenerate horizon.}
	\label{FigCM2}
\end{figure}

\begin{figure}[!h]
	\centering
	\begin{subfigure}[b]{0.5\linewidth}
		\includegraphics[width=\linewidth]{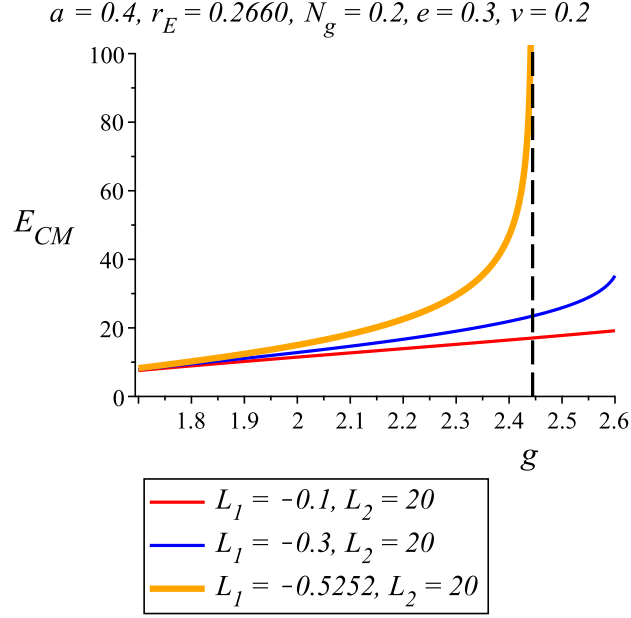}
			\end{subfigure}
	\caption{For an extremal BH case the variation of $E_{CM}$ with $g$ for three combinations of $L_1$ and $L_2=L_c$,  where the vertical black dashed line corresponds to the degenerate horizon.}
	\label{FigCM3}
\end{figure}
\end{itemize}
\newpage
\subsection{Near-Horizon Collision in Nonextremal Spacetime}

\begin{itemize}
	\item Jacobson and Sotiriou pointed out that infinite energies for the colliding particles can only be attained at infinite proper time \cite{52}. Thus this type of collision process does not take place in the real world. However, for the case of a non-extremal BH, the proper time for the particle to reach the horizon is although large but finite. So, it seems worth considering particle collision around a non-extremal BH spacetime. The form of $E_{CM}$  is the same as (\ref{11}) with replacement, $L_c \to {L_c}^\prime$. Here we present some results: the range of angular momentum  $(L_{min}, L_{max})$ for a sample $g$ = 0.1.
	
	\begin{table}[!h]
		\centering
		\begin{tabular}{|c|c|c|c|c|c|c|c|c|}
			\hline 
			\multicolumn{5}{|c|}{$g=0.1$}  \\
			\hline
			{\bf $a$} & {\bf $r_+$} & {\bf $r_-$} & {\bf $L_{max}$} & {\bf $L_{min}$} \\
			\hline
			$0.1$ & $1.906$ & $0.027$ & $36.86$ & -0.5077  \\
			\hline 
			0.3 & 1.859 & 0.08 & 12.59 & -0.7434  \\
			\hline
			0.5 & 1.532 & 0.167 & 7.27 & -0.962  \\
			\hline
		\end{tabular} 
		\caption{Numerical evaluation of maximum and minimum value of angular momentum for the nonextremal BH spacetime for $g$ = 0.1.}
		\label{max}
	\end{table}
	The condition for obtaining a negative effective potential $V_{eff}({L_c}^\prime)$ near the horizon is that the colliding particles should have critical angular  momentum ${L_c}^\prime \in (L_{min}, L_{max})$ as exhibited from the Fig.(\ref{Ang})
	
	\begin{figure}[!h]
		\centering
		\begin{subfigure}[b]{0.375\linewidth}
			\includegraphics[width=\linewidth]{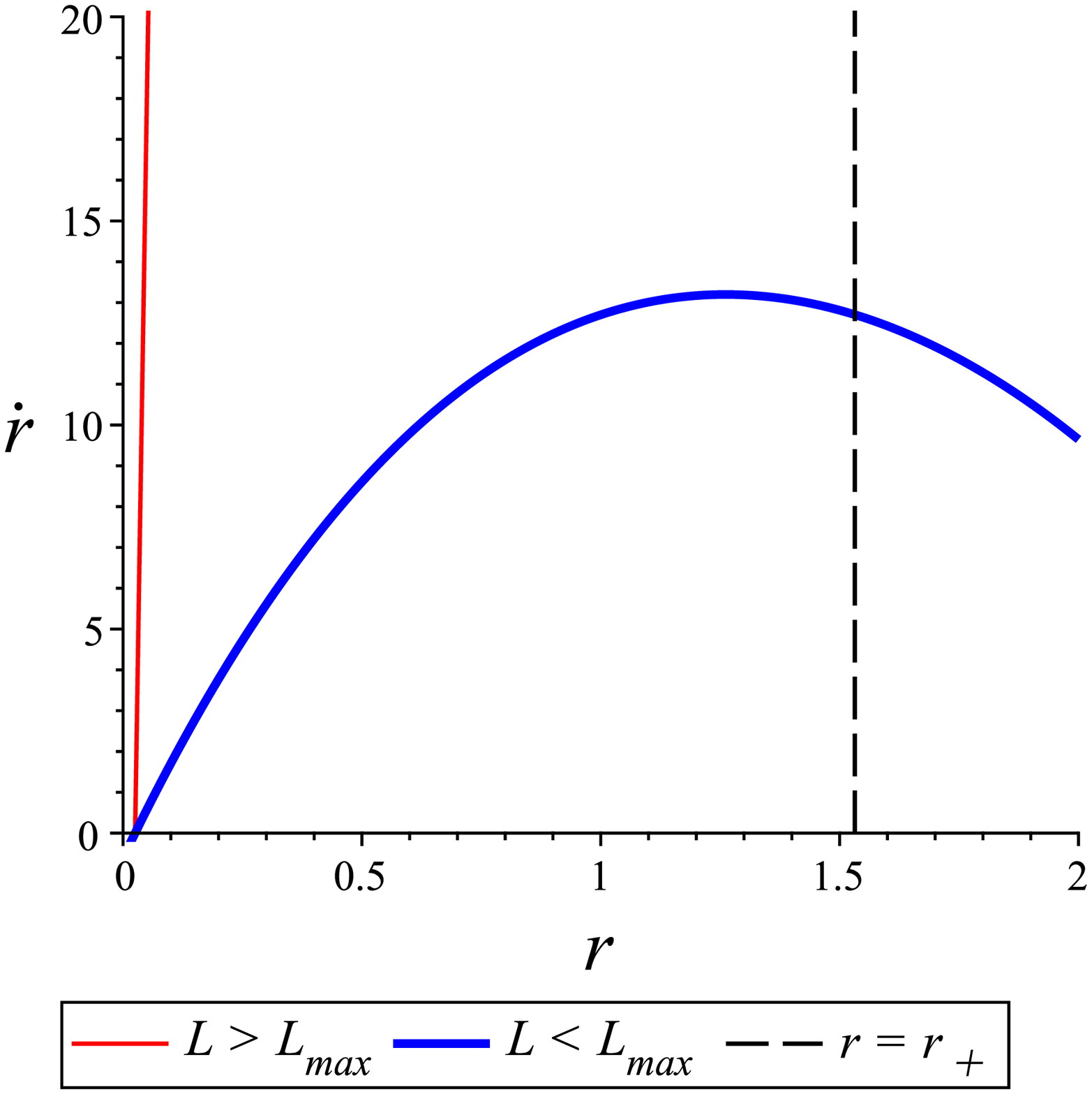}
			\caption{(i)}
		\end{subfigure}
		\begin{subfigure}[b]{0.375\linewidth}
			\includegraphics[width=\linewidth]{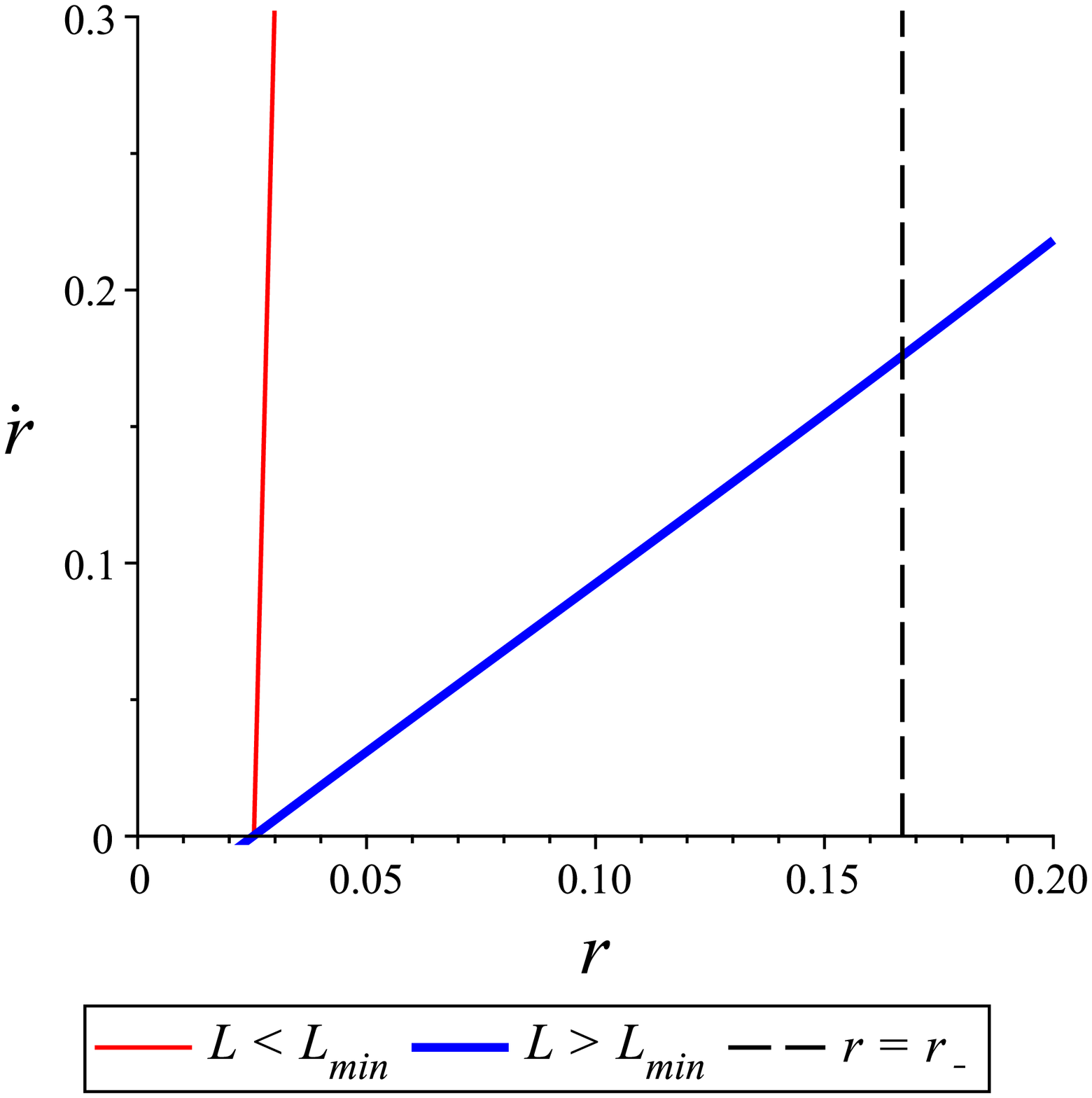}
			\caption{(ii)}
		\end{subfigure}
		\caption{The variation of $\dot{r}$ with $r$ for different values of angular momentum $L$ has been shown for the non-extremal BH with $g = 0.1$, $a = 0.5$. The bold blue line corresponds to the value ${L_c}^\prime$ falling well inside the range $(L_{min}, L_{max})$ to allow the particles to reach the horizon.}
		\label{Ang}
	\end{figure}

    \item A nonextremal BH satisfies the condition $r_+ \ne r_-$ where $r_+$ and $r_-$ denote the outer and the inner horizon respectively. Both the denominator and the numerator in equation (\ref{11}) will become zero as $r \to r_+$. Thanks to the L'H$\hat{\text{o}}$pital's rule again for saving us and revealing what we are looking for. It is prominent that $E_{CM}$ is not divergent for the nonextremal case even if the BH with strong coupling, less spinning or vice versa as illustrated from Fig.(\ref{figCME}).

\begin{figure}[!h]
	\centering
	\begin{subfigure}[b]{0.4\linewidth}
		\includegraphics[width=\linewidth]{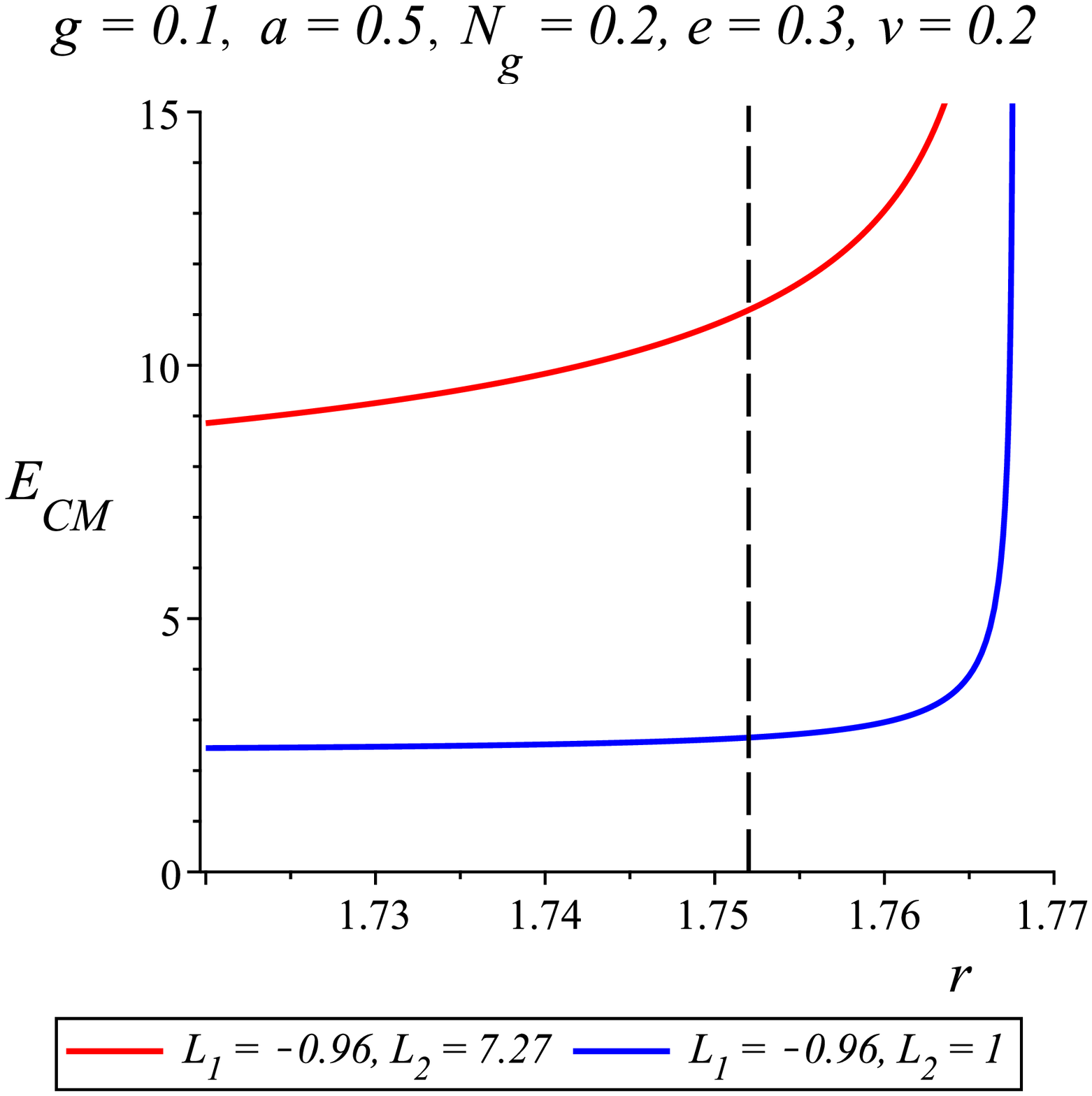}
		\caption{}
	\end{subfigure}
	\begin{subfigure}[b]{0.4\linewidth}
		\includegraphics[width=\linewidth]{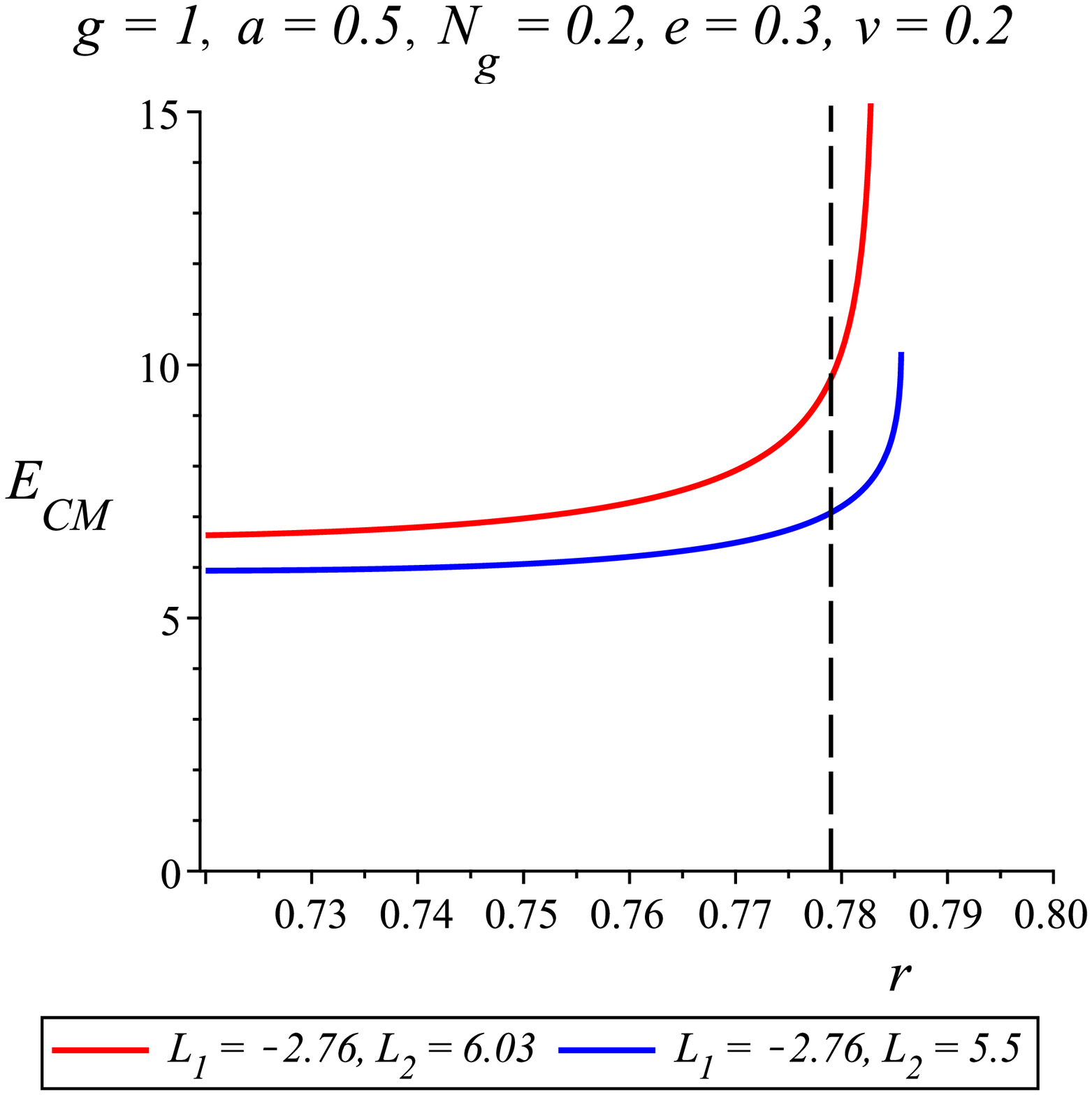}
		\caption{}
	\end{subfigure}
	\caption{For a nonextremal BH case the variation of $E_{CM}$ with $g$ for two combinations of $L_1$ and $L_2$ where the vertical black dashed line corresponds to the event horizon.}
	\label{figCME}
\end{figure}

Furthermore, it seems that initially  $E_{CM}$ decreases with an increase in the value of $g$ and comes to have a lower bound as seen from Fig.(\ref{figECMa}). But $E_{CM}$ approaches an upper bound  if one move slightly beyond $g = 1.2$ which is highlighted in the embedded diagram in Fig.(\ref{figECMa}).

\begin{figure}[h]
\includegraphics[scale=.45]{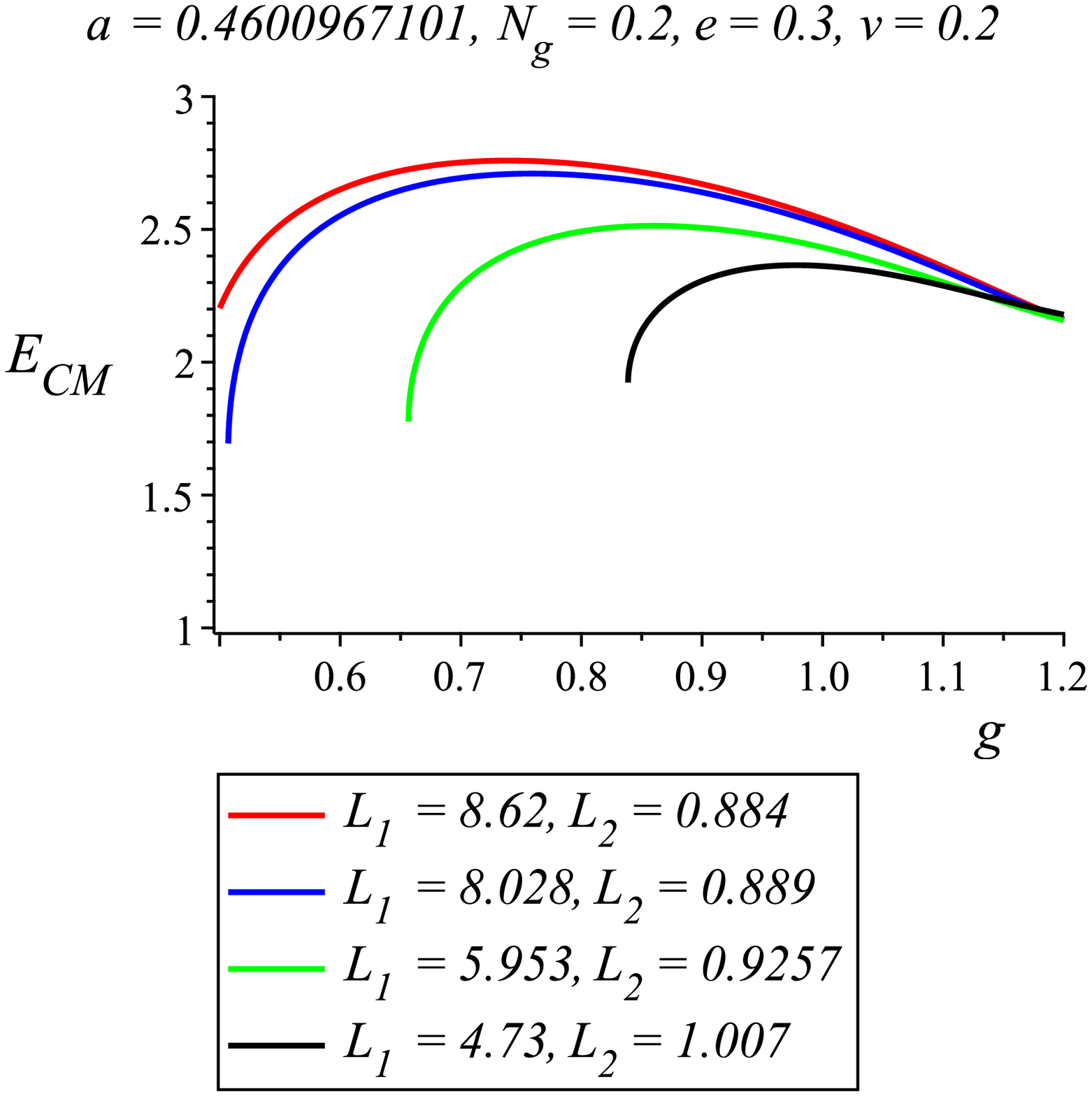}
\centering
\llap{\shortstack{%
        \includegraphics[scale=.135]{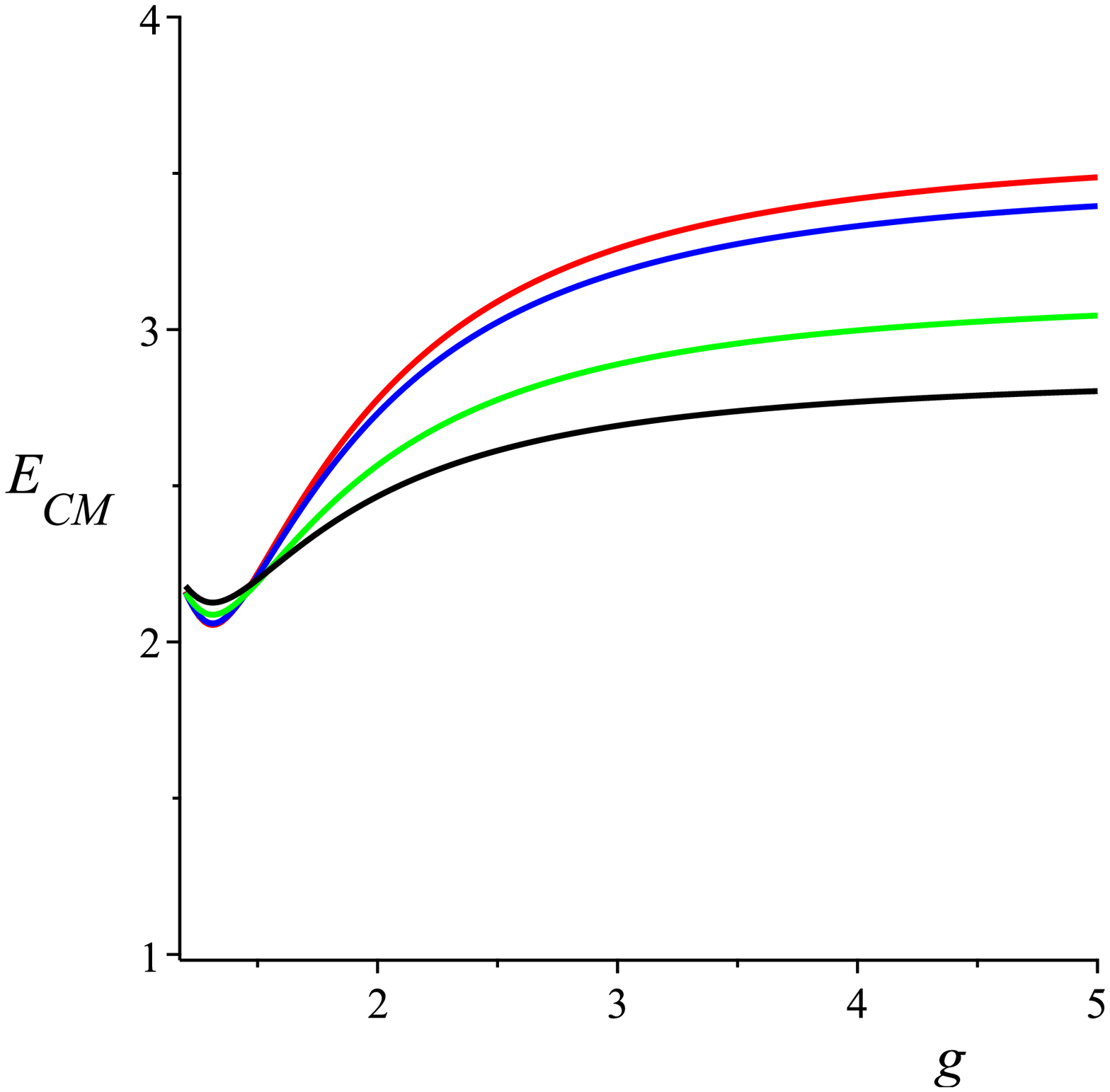}\\
        \rule{0ex}{1.45in}%
      }
 \rule{0.1in}{0ex}}
\caption{The variation of $E_{CM}$ with $g$ for $a=a_E =0.4600967101$. The embedded plot (bottom right) shows the same with slightly extended horizontal range.}
\label{figECMa}
\end{figure}
\end{itemize}
\newpage
\section{ Summary and Conclusions}

The spacetime investigated in this paper are the spacetime of a rotating $U(1)^2$ gauged supergravity BH with dyonic charges. To conclude, we provide below, in a systematic way, a summary of the results obtained.
\begin{itemize}

\item [(i)] The structure of the horizon for this nontrivial spacetime has been reviewed in order to find the critical value of the mass which can be translated to critical (or extreme) values of parameters $a=a_E$ and $g=g_E$ which is the signature of an extremal BH with degenerate horizon.

\item [(ii)] The equations of motions of the energy extraction processes are derived and has been examined accordingly.The most interesting results are observed when $g$ becomes stronger in extremal BH constrained with $N_g=v$. The Penrose process is found to be more efficient ($\sim$ 39 percent) than that is for the KBH in this scenario. Moreover by the Penrose process, we can extract about 60 percent of the initial mass from an extremal BH.

\item [(iii)] In superradiant scattering  for $\omega_m \Omega_H < 0$ the flux of the energy momentum going through the outer horizon turns out to be negative but the flux are positive in infinity. It therefore indeed extracts energy from the BH. Next the influence of the gauge coupling constant $g$ on $\Omega_H$,the angular velocity of the BH at outer horizon have been inquired thoroughly. It is strikingly noticed no superradiance occurs for strong enough coupling.

\item [(iv)] In view of original BSW mechanism fitted for the spacetime (\ref{12}) we estimated $E_{CM}$ for a pair of colliding particles moving near the horizon for both extremal and nonextremal BH. In case of an extremal BH, $E_{CM}$ blows up under some restrictions on the angular momentum. This unleashed nature of $E_{CM}$ can be envisaged to open up a new window to explain physics at the Planck energy scale. On the other hand for nonextremal BH $E_{CM}$ remains finite with a finite upper bound.

\item  [(v)] Even if for relatively less spinning BH, the arbitrarily large center of mass energies can be achieved simply due to the presence of strong gauge coupling constant.

\item [(vi)] We have seen Kerr BH can act as the accelerators of neutral particles. But here in this article we are dealing with dyonic BH which incorporate the magnetic field as well and can therefore act as the accelerators of charged particle to unboundedly high energy. In addition, the supergravity theory, based on the particle symmetry, includes a collection of fields that together have a long-range gravity force with a superpartner. So it is natural to believe gravitational particle acceleration takes significantly extremes to be robust for the BH in supergravity scenario than that is for the Kerr and other generalized Kerr BH in general relativity and other alternative theories of gravity. From such a consideration, rotating dyonic BH in $N=2$ gauged supergravity can be regarded to be more influential super collider ever which further deepen the understanding  of new physics with astrophysical applications e.g. ultra-high energetic dark matter collision at the galactic center, some indirectly observable signatures on the spectra of cosmic rays, neutrinos and gravitational waves in a more obvious way. 

\item [(vii)] The role of $g$ can be summarized in such a way that it provides a reasonable upper bound of the limit of energy extraction process, the possibility of superradiance and how much collision of particles around a supergravity BH may provide a possible detection.

\end{itemize}

\textit {In this paper, we have shown how and to what extent a gauged supergravity BH can be used as a particle accelerator. In principle, it is of no surprise since supersymmetry is the most elegant and experimentally searched for theories in particle physics beyond standard model. But in reality, there are severe astrophysical limitations on the feasibility of such a collision. In addition to that, any beyond SM exotic particles, the BH can produce, would decay quickly. But a recent article \cite{50} suggests that we might be able to observe them indirectly through gravitational waves (GWs) created by the BH. So a BH particle accelerator wouldn’t be nearly as precise as one on Earth, but by looking for fluctuations in the GWs, one could possibly observe that the exotic particles exist. However, at the moment it’s really a great theoretical opportunity to consider a supergravity BH particle accelerator which would be the most worth trying to provide another piece of the dark matter puzzle}.\\

Since the BH is asymptotically AdS, superradiance may induce superradiant instability under some conditions. Future work may therefore be the study under which conditions does superradiant instability occur. Also it would definitely be interesting and meaningful to analyze different aspects of the strong field gravitational lensing and shadows around the gauged supergravity BH spacetime in the future.

\section*{Acknowledgments}
 The authors are indebted  to the anonymous referee for the constructive comments and suggestions which helped us to improve the presentation of this paper. AR is thankful to Oindrila Ganguly and Tomohiro Harada for the fruitful and interesting discussions. H.N. thankfully acknowledges the financial support provided by Science and Engineering Research Board (SERB) during the course of this work through grant number EMR/2017/000339. R.G. is supported by FONDECYT project No 1171384.

\end{document}